\documentclass[journal,10pt]{IEEEtran} 
\usepackage[figurename=Fig.,justification=centering]{caption}
\usepackage{tikz}
\usetikzlibrary{calc, positioning, shapes.geometric, decorations.pathreplacing}
\usepackage{array}
\usepackage{verbatim}
\usetikzlibrary{calc, shapes, arrows, positioning}
\usepackage{authblk}
\usepackage{blindtext}
\usepackage{ifpdf}
\usepackage{graphicx}
\usepackage{tabulary}
\usepackage{amsmath,amsthm,amssymb}

\usepackage{kantlipsum}
\allowdisplaybreaks
\usepackage{cleveref}
\usepackage{lipsum}%
\usepackage{optidef}
\usepackage[english]{babel} 
\theoremstyle{plain}
\newtheorem{theorem}{Theorem}

\crefname{theorem}{Theorem}{theorem}
\crefname{lemma}{Lemma}{Lemmas}

\usepackage{cite}
\usepackage{dsfont}
\usepackage{color}
\usepackage{algorithm}
\usepackage{algorithmicx, algpseudocode}
\usepackage{etoolbox}
\usepackage{subcaption}
\usepackage{balance}
\usepackage{empheq,etoolbox}
\usepackage[utf8]{inputenc}
\usepackage{multirow}
\usepackage{float}

\usepackage{amssymb}
\usepackage{pifont}
\usepackage[hmargin=1.27cm,top=1.3cm,bottom=1.3cm]{geometry}

\usepackage{physics}

\floatstyle{plaintop}
\restylefloat{table}
\usepackage{lipsum} 
\allowdisplaybreaks

\tikzset{brace/.style={decorate, decoration={brace}},
 brace mirrored/.style={decorate, decoration={brace,mirror}},
}

\newcounter{brace}
\newcounter{arrow}
\newif\ifmarkchanges
\ifdefined\CleanVersion\markchangesfalse\else\markchangestrue\fi

\begin{document}
 \captionsetup[figure]{name={Fig.},labelsep=period}


\title{{\color{black}Robust} Joint Power Allocation and Antenna Placement for Pinching-Antenna Systems under {\color{black}Gaussian} Location Uncertainty}

\author{Hao Feng, Ming Zeng, \textit{Senior Member, IEEE}, Xingwang Li, \textit{Senior Member, IEEE},  Wenwu Xie and Nian Xia
    \thanks{Copyright (c) 20xx IEEE. Personal use of this material is permitted. However, permission to use this material for any other purposes must be obtained from the IEEE by sending a request to pubs-permissions@ieee.org.}
    \thanks{The work of M. Zeng and H. Feng were supported by NSERC of Canada through the Discovery Grant, and the China Scholarship Council (CSC, No. 202506630002), respectively. (Corresponding author: Ming Zeng.)}
    \thanks{H. Feng is with Hunan Institute of Engineering, Xiangtan, China, and also with Donghua University, Shanghai, China, (e-mail: 1219001@mail.dhu.edu.cn).}
    
    \thanks{H. Feng and M. Zeng are with Laval University, Quebec City, Canada (email: ming.zeng@gel.ulaval.ca).}

    \thanks{X. Li is with Henan Polytechnic University, Jiaozuo, China (email: lixingwang@hpu.edu.cn).}

    \thanks{W. Xie is with Hunan Institute of Science and Technology, Yueyang, China (e-mail: gavinxie@hnist.edu.cn).}



    \thanks{{N. Xia is with Nanjing Normal University, Nanjing 210023, China. e-mail: nian.xia@nnu.edu.cn}} 

    











    }
\maketitle

\begin{abstract}
Pinching antenna systems (PASS) have attracted much attention recently owing to their capability to maintain reliable line-of-sight communication in high-frequency bands. By guiding signals through a waveguide and emitting them via a movable pinching antenna, PASS enable dynamic control of signal propagation and spatial adaptability. However, their performance heavily depends on effective resource allocation, including power allocation and antenna positioning, which becomes challenging under user localization uncertainty. Existing studies largely assume perfect user locations, while our prior work considered uniformly distributed localization errors. In this paper, we develop a robust resource allocation framework for multi-user downlink PASS under Gaussian-distributed localization uncertainty. An energy-efficiency (EE) maximization problem is formulated subject to probabilistic outage constraints, and the minimum transmit power is analytically characterized for any given antenna position. For the unbounded Gaussian model, we prove that the resulting one-dimensional antenna-placement objective is convex and derive a globally optimal derivative-bisection algorithm. {\color{black}We further consider a circularly truncated Gaussian model for bounded localization-error regions and develop a deterministic branch-and-bound method that returns an $\epsilon_{\rm B}$-globally optimal antenna position. Finally, we discuss how position-dependent waveguide attenuation modifies the placement objective and may destroy its convexity. Simulation results verify that the proposed branch-and-bound solution closely matches the bounded grid-search benchmark, demonstrate substantial EE gains from optimized pinching-antenna placement over the bounded fixed-antenna architecture, and reveal an increasing distinction between the bounded and unbounded Gaussian models as the location uncertainty grows.}
\end{abstract}

\IEEEpeerreviewmaketitle

\section{Introduction} 

Pinching antenna systems have recently attracted considerable interest as a potential architecture for restoring line-of-sight (LoS) communication links in high-frequency regimes, including millimeter-wave (mmWave) and terahertz (THz) bands \cite{Atsushi_22, yang2025,  ding2024, pass1, Xiao25, wijewardhana2025}, and are therefore regarded as a key enabler for 6G and beyond networks. In such systems, user signals are first transmitted through a dielectric waveguide and subsequently radiated by a pinching antenna attached to it. By dynamically adjusting the antenna's position along the waveguide, pinching antenna systems provide exceptional flexibility in controlling signal propagation, leading to improved spectral efficiency and adaptability \cite{ xie2025, wang2024, hu2025, Li25, Zeng_COMML25, fu2025, Tegos_2025, Zhao_TCOM25}.

A core challenge in pinching-antenna systems is resource allocation, which involves the joint optimization of transmission parameters such as power, bandwidth, and antenna position to maximize system performance \cite{zeng2025_WCM}. Most existing studies assume perfect channel state information (CSI) or accurate user location knowledge. However, in realistic deployments, channel estimation errors and localization inaccuracies are inevitable due to noise, hardware impairments, environmental dynamics, and pilot overhead limitations \cite{Xiao_COMML25}. Ignoring such uncertainties can lead to substantial performance degradation, underscoring the necessity of robust resource allocation designs.

To the best of the authors' knowledge, only two prior works have addressed robust resource allocation in pinching-antenna systems \cite{Sun_TVT26, zeng2025robust}. In \cite{Sun_TVT26}, a robust beamforming design is developed under imperfect CSI, where the actual channel is modeled as the superposition of an estimated component and an additive estimation error. However, the wireless channel component associated with the estimated CSI is assumed to be fixed and independent of the pinching-antenna position, which limits the practical applicability of the model, as the channel inherently depends on antenna placement. 
To overcome this limitation, \cite{zeng2025robust} considers a joint robust power allocation and antenna position optimization framework under user location uncertainty. 
{The uniform-error model in \cite{zeng2025robust} provides a bounded uncertainty description with constant density. A complementary Gaussian model captures uncertainty concentrated around an available position estimate. Such models have been used in spatial wireless channel prediction \cite{Muppirisetty2016}, location-based rate selection and reliability analysis \cite{Kallehauge2024}, and robust near-field beamfocusing \cite{Uchimura2026}. In particular, \cite{Kallehauge2024} motivates a Gaussian localization-estimator model through asymptotic maximum-likelihood theory. This provides an estimation-based rationale for a local Gaussian approximation in a sufficiently accurate, approximately unbiased positioning regime.} 

{A potential vehicular application is infrastructure-to-vehicle downlink access in a parking facility or vehicle depot equipped with a fixed waveguide. Position-assisted transmission for vehicular links has been investigated, for example, in \cite{LiuVehicular2020}. In the considered application, the access point (AP) uses estimates of vehicle-terminal positions to jointly select the pinching-antenna position and transmit powers while meeting prescribed rate and outage requirements. This connects the proposed framework to reliable, energy-efficient vehicular access under imperfect location information.}

{\color{black}Motivated by these observations, this paper investigates robust resource allocation for pinching-antenna systems under stochastic Gaussian user-location uncertainty, considering both unbounded and circularly truncated Gaussian models.}
Specifically, we formulate an EE maximization problem for a multi-user downlink pinching-antenna system subject to user-specific probabilistic outage constraints. We first derive the minimum transmit power required by each user for any fixed pinching-antenna position. For the unbounded Gaussian model, we then establish the convexity of the remaining one-dimensional placement problem and obtain its globally optimal solution through bisection on the analytical objective derivative. {\color{black}To account for physically bounded localization errors, we further formulate a circularly truncated Gaussian model and develop a deterministic branch-and-bound placement algorithm with an $\epsilon_{\rm B}$-global optimality guarantee. Finally, we discuss how position-dependent waveguide attenuation changes the placement objective and the associated optimality properties. Numerical results validate the bounded branch-and-bound design against grid search, quantify the EE gain provided by antenna-position optimization, and illustrate the impact of truncating the Gaussian location-error distribution.}

\section{System Model and Problem Formulation}
\subsection{System Model}
We examine a downlink multi-user scenario where an AP, located at $(0,0,0)$ meters communicates with $N$ single-antenna users with the aid of a pinching antenna mounted on a waveguide. 
The waveguide is placed following the $x$-axis, with a length of $L$ meters and a height of $d$ meters. 

Denote the estimated location of user $n, n \in \{1, \dots, N\}$ as $\hat{\Phi}_n = (\hat{x}_n, \hat{y}_n, 0)$ meters. Unlike conventional studies assuming perfect location knowledge, we account for location uncertainty, {that is, user $n$ is located at $\Phi_n=(x_n,y_n,0)$, and, conditioned on the available estimate, its horizontal coordinates follow an isotropic normal distribution $(X_n,Y_n) \sim \mathcal{N}({\boldsymbol{\mu}}, {\boldsymbol{\Sigma}})$}, where the mean vector is ${\boldsymbol{\mu}} = (\hat{x}_n, \hat{y}_n)$, and the covariance matrix is ${\boldsymbol{\Sigma}} = \sigma^2 {\bf{I}} = \begin{pmatrix} \sigma^2 & 0 \\ 0 & \sigma^2 \end{pmatrix}$. A higher variance value $\sigma^2 $ means higher positioning error and vice versa. 
{The Gaussian model captures residual uncertainty around an estimated position and is motivated by the asymptotic Gaussian behavior of localization estimators under suitable conditions \cite{Kallehauge2024}. 
The isotropic covariance assumes equal horizontal error variances and negligible cross-correlation, while a common \(\sigma\) assumes comparable positioning accuracy across users; these assumptions should be validated against positioning-error statistics in the intended deployment, and Section IV considers a circularly truncated model when bounded errors are assumed.}
The users are scheduled using time-division multiple access (TDMA) to avoid mutual interference \cite{ding2024, xie2025}. To avoid frequent repositioning across TDMA slots, the users share a common pinching antenna location \cite{Tegos_2025}, expressed as $\Phi^{\text{Pin}} = (x^{\text{Pin}}, 0, d)$ meters, where $x^{\text{Pin}} \in [0, L]$. 
{\color{black}This single-waveguide, common-position architecture limits the available spatial degrees of freedom. Extending the proposed framework to multi-waveguide architectures with multiple independently adjustable pinching-antenna positions is an important direction for future work. In practical implementations, the feasible pinching-antenna locations may be restricted to a finite set \cite{Tyrovolas_IoTJ26}; the continuous-position model considered here therefore serves as an ideal benchmark, while the same per-position power allocation can be directly evaluated over any prescribed discrete set.}
\begin{equation} \label{rate_expression}
    R_n=\log_2 \left(1+ \frac{  \eta P_n }{ \abs{ {\Phi}_n - \Phi^{\text{Pin}}}^2 \sigma_n^2} \right). 
\end{equation}
Here, $P_n$ is the transmit power for user $n$, while $\sigma_n^2$ represents the corresponding noise power. $\eta=\frac{c^2}{16 \pi^2 f^2}$, $c$ is the speed of light, and $f$ is the carrier frequency \cite{Tegos_2025}.

Owing to positioning error in user locations, {user $n$'s horizontal position $(X_n,Y_n)$} is modeled as a bivariate Gaussian random vector, which in turn makes the instantaneous achievable rate $R_n$ a random variable. A transmission outage occurs whenever the instantaneous achievable rate $R_n$ drops below a predefined target $\hat{R}_n$ for user $n$. The outage probability is expressed as $\Pr\left[ R_n < \hat{R}_n \,\middle|\, \hat{\Phi}_n, \sigma^2 \right]$,
denoting the likelihood that $R_n$ is less than the required target $\hat{R}_n$, conditioned over the estimated position $\hat{\Phi}_n$ and the error variance $\sigma^2 $. This metric characterizes the reliability of the AP---user $n$ link under positioning error.

\subsection{Problem Formulation}
To ensure the quality of service (QoS), the outage probability for user $n$, $\forall n$ cannot exceed a maximum allowable threshold $\varepsilon_n$, i.e., $\Pr\left[ R_n < \hat{R}_n \,\middle|\, \hat{\Phi}_n, \sigma^2 \right] \leq \varepsilon_n,  \forall n.$

Under this constraint, we aim to maximize the system's EE, which is defined as the ratio of the effective sum rate, i.e., $\sum_{n=1}^N (1- \varepsilon_n) \hat{R}_n$ \cite{Fang_TCOM21} over the total transmit power, i.e., $\sum_{n=1}^N P_n $ \cite{ Zhang_TVT17, Zeng_TVT19}. 
{\color{black}A position-independent circuit power term could be included without affecting the optimal power allocation or antenna placement, and is therefore omitted to focus on the transmit-power impact of localization uncertainty.}
The problem is formulated as
\begin{subequations} \label{P_NOMA}
   \begin{align}
       \max_{x^{\text{Pin}}, P_n}~ & \frac{\sum_{n=1}^N (1- \varepsilon_n) \hat{R}_n }{\sum_{n=1}^N P_n} \\
       \text{s.t.}~~ & x^{\text{Pin}} \in [0, L], \label{P_NOMA_b} \\
       & \Pr\left[ R_n < \hat{R}_n \,\middle|\, \hat{\Phi}_n, \sigma^2 \right] \leq \varepsilon_n, \quad \forall n, 
   \end{align}
\end{subequations}
where (\ref{P_NOMA}b) limits the pinching antenna to the range of the waveguide, while (\ref{P_NOMA}c) constrains the outage probability by $\varepsilon_n$.


\section{Proposed Solution}
Since the parameters $\varepsilon_n$ and $\hat{R}_n$ are prescribed QoS requirements, $\sum_{n=1}^N(1-\varepsilon_n)\hat{R}_n$ is a constant. Therefore, the above EE maximization problem can be equivalently transformed into the following total power minimization problem. 
\begin{equation} \label{P_NOMA_2}
    \min_{x^{\text{Pin}}, P_n}~ \sum_{n=1}^N P_n 
       \text{s.t.}~ (\ref{P_NOMA}\text{b}), (\ref{P_NOMA}\text{c})
\end{equation}

In problem~\eqref{P_NOMA_2}, both the objective function and constraint (\ref{P_NOMA}b) are linear. The major difficulty of solving it arises from the probabilistic outage constraints in (\ref{P_NOMA}c), which are non-convex. Furthermore, the pinching antenna position enters the outage constraints of all users, thereby inducing coupling among these constraints. Consequently, they cannot be decoupled and must be addressed jointly, which significantly increases the overall problem complexity.
To address problem~\eqref{P_NOMA_2}, we first consider the case where the pinching antenna position is given. Based on the insight gained from this case, we will handle the pinching antenna position optimization in the end. 

\subsection{Power Optimization under given Pinching Antenna Position}

For a given pinching-antenna position, the users' transmit powers are decoupled. Hence, problem~\eqref{P_NOMA_2} reduces to $N$ independent single-user subproblems:
\begin{subequations} \label{P_NOMA_4}
   \begin{align}
       \min_{ P_n}~ &  P_n \\
       \text{s.t.}~~ 
       & \Pr\left[ R_n < \hat{R}_n \,\middle|\, \hat{\Phi}_n, \sigma^2 \right] \leq \varepsilon_n,  
   \end{align}
\end{subequations}
We now examine constraint~(\ref{P_NOMA_4}b). For a given pinching-antenna position and transmit power, the achievable rate in~(1) depends only on the three-dimensional distance between the user and the pinching antenna. Let $R$ denote the distance at which the target rate $\hat{R}_n$ is achieved. The intersection of the corresponding sphere centered at $\boldsymbol{\Phi}^{\mathrm{Pin}}$ with the user
plane $z=0$ forms a coverage disk centered at
$(x^{\mathrm{Pin}},0)$ with radius $r=\sqrt{R^2-d^2}$. Hence, a user location inside this disk satisfies
$R_n\geq\hat{R}_n$. The following theorem characterizes the
corresponding coverage probability.

\begin{theorem}\label{theorm 1}
The probability that the true user location lies within the
coverage disk of radius $r$ is given by
\begin{equation}
\Pr\left\{
(X_n-x^{\mathrm{Pin}})^2+Y_n^2\leq r^2
\right\}
=
1-Q_1\left(
\frac{l}{\sigma},
\frac{r}{\sigma}
\right),
\end{equation}
where
\begin{equation*}
Q_1(a,b)
=
\int_b^\infty
x\exp\left(-\frac{x^2+a^2}{2}\right)I_0(ax)\,dx
\end{equation*}
is the first-order Marcum-$Q$ function, and $I_0(\cdot)$ is the
modified Bessel function of the first kind and order zero. Moreover,
\begin{equation*}
l=
\sqrt{
(\hat{x}_n-x^{\mathrm{Pin}})^2+\hat{y}_n^2
}
\end{equation*}
denotes the distance between the projected antenna position
$(x^{\mathrm{Pin}},0)$ and the estimated location of user $n$.
\end{theorem}

\begin{IEEEproof}
Define the normalized random variables
\begin{equation*}
W_1=\frac{X_n-x^{\mathrm{Pin}}}{\sigma},
\qquad
W_2=\frac{Y_n}{\sigma}.
\end{equation*}
Since $X_n$ and $Y_n$ are independent Gaussian random variables,
$W_1$ and $W_2$ are independent and satisfy
\begin{equation*}
W_1\sim\mathcal{N}\left(
\frac{\hat{x}_n-x^{\mathrm{Pin}}}{\sigma},1
\right),
\qquad
W_2\sim\mathcal{N}\left(
\frac{\hat{y}_n}{\sigma},1
\right).
\end{equation*}
Consequently,
\begin{equation*}
S=W_1^2+W_2^2
\sim
\chi'^2_2\left(\lambda\right),
\qquad
\lambda=
\frac{
(\hat{x}_n-x^{\mathrm{Pin}})^2+\hat{y}_n^2
}{\sigma^2}
=
\frac{l^2}{\sigma^2},
\end{equation*}
where $\chi'^2_2(\lambda)$ denotes a noncentral chi-squared
distribution with two degrees of freedom and noncentrality parameter
$\lambda$. Therefore,
\begin{align}
&\Pr\left\{
(X_n-x^{\mathrm{Pin}})^2+Y_n^2\leq r^2
\right\} =
\Pr\left\{
S\leq\frac{r^2}{\sigma^2}
\right\} \notag\\
&\quad=
1-Q_1\left(
\sqrt{\lambda},
\frac{r}{\sigma}
\right)
=
1-Q_1\left(
\frac{l}{\sigma},
\frac{r}{\sigma}
\right),
\end{align}
where the second equality follows from the CDF of a noncentral
chi-squared random variable with two degrees of freedom
\cite{Marvin}.
\end{IEEEproof}

To satisfy the outage constraint, we have 
\begin{equation} \label{r constraint}
    1-  Q_1\left(\frac{l}{\sigma}, \frac{r}{\sigma}\right) \geq 1- \varepsilon_n \rightarrow Q_1\left(\frac{l}{\sigma}, \frac{r}{\sigma}\right) \leq \varepsilon_n.
\end{equation}

Since $Q_1(l/\sigma,r/\sigma)$ is decreasing in $r$, the minimum coverage radius $r_{\min}$ is uniquely determined by
$Q_1\!\left(\frac{l}{\sigma},\frac{r_{\min}}{\sigma}\right)
=\varepsilon_n$, and can be efficiently obtained by bisection.


At $r=r_{\min}$, the minimum power satisfying the outage constraint is
\begin{equation}
P_n^{\min}
=
\frac{(2^{\hat R_n}-1)(r_{\min}^2+d^2)\sigma_n^2}{\eta}.
\label{Pmin}
\end{equation}


{\color{black}
\subsection{Globally Optimal Pinching-Antenna Placement}
For notational convenience, define
\begin{equation}
    c_n=\frac{(2^{\hat{R}_n}-1)\sigma_n^2}{\eta}
\end{equation}
and, for a candidate antenna position $x\in[0,L]$,
\begin{equation}
    l_n(x)=\sqrt{(x-\hat{x}_n)^2+\hat{y}_n^2}.
    \label{eq:ln_x}
\end{equation}
Let $r_{n,\min}(x)$ denote the unique radius satisfying
\begin{equation}
    Q_1\left(\frac{l_n(x)}{\sigma},\frac{r_{n,\min}(x)}{\sigma}\right)=\varepsilon_n.
    \label{eq:rmin_implicit}
\end{equation}
Using \eqref{Pmin}, the remaining antenna-placement problem is
\begin{equation}
    \min_{0\leq x\leq L} F_{\rm G}(x),
    \qquad
    F_{\rm G}(x)=\sum_{n=1}^{N}c_n\left(r_{n,\min}^2(x)+d^2\right).
    \label{eq:placement_gaussian}
\end{equation}
Although $r_{n,\min}(x)$ is obtained numerically from \eqref{eq:rmin_implicit}, it is differentiable with respect to $l_n$. In particular, the standard derivatives of the first-order Marcum-$Q$ function are \cite{Pratt_IEEEProc68}
\begin{subequations}
\begin{align}
    \frac{\partial Q_1(a,b)}{\partial a}
    &=b\exp\left(-\frac{a^2+b^2}{2}\right)I_1(ab),\\
    \frac{\partial Q_1(a,b)}{\partial b}
    &=-b\exp\left(-\frac{a^2+b^2}{2}\right)I_0(ab),
\end{align}
\label{eq:marcum_derivatives}
\end{subequations}
where $I_1(\cdot)$ is the modified Bessel function of the first kind and order~1. Applying implicit differentiation to \eqref{eq:rmin_implicit} yields
\begin{equation}
    \frac{d r_{n,\min}}{d l_n}
    =\rho(z_n)
    \triangleq
    \frac{I_1(z_n)}{I_0(z_n)},
    \qquad
    z_n=\frac{l_n r_{n,\min}}{\sigma^2}.
    \label{eq:rmin_derivative}
\end{equation}
The derivative in \eqref{eq:rmin_derivative} is understood by continuity at $l_n=0$, for which $\rho(0)=0$.

\begin{theorem}
The objective function $F_{\rm G}(x)$ in \eqref{eq:placement_gaussian} is convex on $[0,L]$. Consequently, any point satisfying the first-order optimality condition is globally optimal.
\label{theorem:convex_gaussian}
\end{theorem}
\begin{IEEEproof}
The ratio $\rho(z)=I_1(z)/I_0(z)$ is nonnegative and monotonically increasing for $z\geq0$ \cite{segura2021monotonicity}. From \eqref{eq:rmin_derivative},
\begin{equation}
    \frac{d^2r_{n,\min}}{dl_n^2}
    =\rho'(z_n)
    \frac{r_{n,\min}+l_n\rho(z_n)}{\sigma^2}
    \geq0.
\end{equation}
Hence, $r_{n,\min}(l_n)$ is nonnegative, nondecreasing, and convex in $l_n$. It follows that $r_{n,\min}^2(l_n)+d^2$ is also nondecreasing and convex in $l_n$. Moreover, $l_n(x)$ in \eqref{eq:ln_x} is convex in $x$. By the composition rule for a convex nondecreasing function and a convex function, every term $c_n(r_{n,\min}^2(l_n(x))+d^2)$ is convex in $x$. Their sum $F_{\rm G}(x)$ is therefore convex.
\end{IEEEproof}

Using \eqref{eq:rmin_derivative} and
\begin{equation}
    \frac{d l_n(x)}{dx}=\frac{x-\hat{x}_n}{l_n(x)},
\end{equation}
the derivative of the objective is
\begin{equation}
\begin{split}
    F_{\rm G}'(x)
    =\sum_{n=1}^{N}
    2c_n r_{n,\min}(x)
    \rho\left(z_n(x)\right)
    \frac{x-\hat{x}_n}{l_n(x)}.
\end{split}
\label{eq:F_derivative}
\end{equation}
When $l_n(x)=0$, the corresponding summand in \eqref{eq:F_derivative} is set to its continuous limit, which equals zero. Since $F_{\rm G}(x)$ is convex, $F_{\rm G}'(x)$ is monotonically nondecreasing. Therefore, the globally optimal antenna position is characterized by
\begin{equation}
    x^{\text{Pin},\star}
    =
    \begin{cases}
    0, & F_{\rm G}'(0)\geq0,\\
    L, & F_{\rm G}'(L)\leq0,\\
    \{x\in(0,L):F_{\rm G}'(x)=0\}, & \text{otherwise}.
    \end{cases}
    \label{eq:global_solution_gaussian}
\end{equation}
The optimal solution can be obtained by bisection on $F_{\rm G}'(x)$. At each trial position, $r_{n,\min}(x)$ is first computed by bisection from \eqref{eq:rmin_implicit}. 
If this inner radius bisection uses an upper bound $r_{\rm ub}$, the computational complexity of the nested-bisection procedure is on the order of
\begin{equation}
    \mathcal{O}\left(
    N\log\frac{L}{\epsilon_x}
    \log\frac{r_{\rm ub}}{\epsilon_r}
    \right),
\end{equation}
where $\epsilon_x >0$ and $\epsilon_r>0$ denote the prescribed accuracy of the outer and inner bisection search, respectively.


\subsection{Impact of Waveguide Attenuation on Antenna Placement}
The preceding derivations neglect position-dependent attenuation along the waveguide. To briefly characterize its effect, denote the power-attenuation coefficient in dB/m by $\alpha_{\rm wg}$. Then, the power-transfer factor becomes \cite{tyrovolas2025}
\begin{equation}
    \xi_{\rm wg}(x)=10^{-\alpha_{\rm wg}x/10}=e^{-\kappa x},
    \qquad
    \kappa=\frac{\ln 10}{10}\alpha_{\rm wg}.
\end{equation}
If $P_n$ denotes the power injected into the waveguide, the minimum required input power is multiplied by $e^{\kappa x}$. Hence, for the unbounded Gaussian model, the placement objective becomes
\begin{equation}
    F_{\rm wg}(x)=e^{\kappa x}F_{\rm G}(x),
    \label{eq:wg_objective}
\end{equation}
with
\begin{subequations}
\begin{align}
    F_{\rm wg}'(x)
    &=e^{\kappa x}\left(F_{\rm G}'(x)+\kappa F_{\rm G}(x)\right),\\
    F_{\rm wg}''(x)
    &=e^{\kappa x}\left(F_{\rm G}''(x)+2\kappa F_{\rm G}'(x)+\kappa^2F_{\rm G}(x)\right).
\end{align}
\label{eq:wg_derivatives}
\end{subequations}
Although $F_{\rm G}(x)$ is convex, the attenuated objective
$F_{\rm wg}(x)=e^{\kappa x}F_{\rm G}(x)$ is not necessarily
convex. Convexity is preserved if and only if $ F_{\rm wg}''(x) \geq 0, 
\forall x\in[0,L]$. 
Under this condition, derivative bisection remains globally
optimal. Otherwise, the attenuated placement problem may become nonconvex and can be solved using the one-dimensional
branch-and-bound framework introduced in the following section.
}

{\color{black}
\section{Truncated Gaussian Location Uncertainty}
The Gaussian location uncertainty model considered above has unbounded support. In practice, however, the true user location may be restricted to a bounded uncertainty region around the estimated location. To capture this effect while preserving the local isotropic nature of positioning errors, we consider a circularly truncated Gaussian uncertainty model.

Define the bounded circular uncertainty region as
\begin{equation}
    \mathcal{U}_n
    =
    \left\{
    (x,y):
    (x-\hat{x}_n)^2+(y-\hat{y}_n)^2\leq r_{\rm u}^2
    \right\},
\end{equation}
where $r_{\rm u}$ is the radius of the bounded localization-error region. Under the circularly truncated Gaussian model, the true location follows the original isotropic Gaussian distribution conditioned on lying inside $\mathcal{U}_n$. Hence, its probability density function is
\begin{equation}
    f_n^{\rm TG}(x,y)
    =
    \frac{f_n^{\rm G}(x,y)}{Z_n},
    \quad (x,y)\in\mathcal{U}_n,
\end{equation}
and $f_n^{\rm TG}(x,y)=0$ otherwise. Here,
\begin{equation}
    f_n^{\rm G}(x,y)
    =
    \frac{1}{2\pi\sigma^2}
    \exp\left(
    -\frac{(x-\hat{x}_n)^2+(y-\hat{y}_n)^2}{2\sigma^2}
    \right)
\end{equation}
is the original Gaussian density, and
\begin{equation}
    Z_n
    =
    1-\exp\left(-\frac{r_{\rm u}^2}{2\sigma^2}\right)
    \label{eq:tg_normalization}
\end{equation}
is the normalization factor.
{Here, $\sigma$ denotes the standard deviation of the original Gaussian coordinates, rather than the coordinate standard deviation after truncation.}

For a given pinching-antenna position $x^{\text{Pin}}$, define the coverage circle on the user plane as
\begin{equation}
    \mathcal{C}(r)
    =
    \left\{
    (x,y):
    (x-x^{\text{Pin}})^2+y^2\leq r^2
    \right\}.
\end{equation}
The corresponding conditional coverage probability is
\begin{equation}
    \Pi_n^{\rm TG}(r;x^{\text{Pin}})
    =
    \frac{
    \displaystyle
    \int_{\mathcal{U}_n\cap\mathcal{C}(r)}
    f_n^{\rm G}(x,y)\,dxdy
    }
    {Z_n}.
    \label{eq:tg_probability_general}
\end{equation}
Let
\begin{equation}
    l_n(x^{\text{Pin}})
    =
    \sqrt{(\hat{x}_n-x^{\text{Pin}})^2+\hat{y}_n^2}.
\end{equation}

Using global polar coordinates centered at
$(\hat{x}_n,\hat{y}_n)$, a candidate user location is
\begin{equation}
(x,y)
=
(\hat{x}_n,\hat{y}_n)
+
s(\cos\phi,\sin\phi),
\qquad
0\leq s\leq r_{\rm u},
\end{equation}
where $\phi$ is measured from the global $x$-axis. Let
$\alpha_n$ denote the direction from
$(\hat{x}_n,\hat{y}_n)$ toward the projected antenna and
define the relative angle
\begin{equation}
\theta=\phi-\alpha_n.
\end{equation}
The distance from the candidate location to the projected antenna
is then
\begin{equation}
\sqrt{s^2+l_n^2-2sl_n\cos\theta}.
\end{equation}

Define $\Omega(s;l_n,r)$ as the angular measure for which this distance does not exceed $r$. For $l_n>0$,
\begin{equation}
\Omega(s;l_n,r)
=
\begin{cases}
2\pi, & s+l_n\leq r,\\
0, & r\leq |s-l_n|,\\
2\cos^{-1}\left(
\dfrac{s^2+l_n^2-r^2}{2sl_n}
\right), & |s-l_n|<r<s+l_n.
\end{cases}
\label{eq:angular_measure_tg}
\end{equation}
For $l_n=0$, $\Omega(s;0,r)=2\pi$ if $s\leq r$ and $\Omega(s;0,r)=0$ otherwise. Thus,
\begin{equation}
    \Pi_n^{\rm TG}(r;x^{\text{Pin}})
    =
    \frac{
    \displaystyle
    \int_0^{r_{\rm u}}
    \frac{s}{2\pi\sigma^2}
    \exp\left(-\frac{s^2}{2\sigma^2}\right)
    \Omega\left(s;l_n(x^{\text{Pin}}),r\right)\,ds
    }
    {
    \displaystyle
    1-\exp\left(-\frac{r_{\rm u}^2}{2\sigma^2}\right)
    }.
    \label{eq:tg_probability_polar}
\end{equation}
The minimum coverage radius is
\begin{equation}
    r_{n,\min}^{\rm TG}(x^{\text{Pin}})
    =
    \inf\left\{
    r\geq0:
    \Pi_n^{\rm TG}(r;x^{\text{Pin}})\geq1-\varepsilon_n
    \right\}.
    \label{eq:rmin_tg}
\end{equation}
Because \eqref{eq:tg_probability_polar} is nondecreasing in $r$, this value can be computed by bisection over the finite interval
\begin{equation}
    0\leq r\leq l_n(x^{\text{Pin}})+r_{\rm u},
\end{equation}
whose upper endpoint covers the entire truncated uncertainty disk. The corresponding minimum power is
\begin{equation}
    P_{n,{\rm TG}}^{\min}(x^{\text{Pin}})
    =
    c_n\left(\left(r_{n,\min}^{\rm TG}(x^{\text{Pin}})\right)^2+d^2\right),
    \label{eq:Pmin_tg}
\end{equation}
and the bounded-Gaussian placement problem is
\begin{equation}
    \min_{0\leq x\leq L}F_{\rm TG}(x),
    \qquad
    F_{\rm TG}(x)=\sum_{n=1}^{N}P_{n,{\rm TG}}^{\min}(x).
    \label{eq:placement_tg}
\end{equation}

Unlike the unbounded Gaussian case, a convenient closed-form derivative and a general convexity proof for \eqref{eq:placement_tg} are not available because the coverage probability is evaluated through the disk-intersection integral in \eqref{eq:tg_probability_polar}. Nevertheless, the problem remains one-dimensional and can be solved globally through deterministic branch-and-bound. To construct a valid interval lower bound, consider an interval $\mathcal{I}=[a,b]\subseteq[0,L]$ and define
\begin{equation}
    \delta_n(\mathcal{I})
    =
    \begin{cases}
    0, & \hat{x}_n\in[a,b],\\
    \min\{|a-\hat{x}_n|,|b-\hat{x}_n|\}, & \text{otherwise},
    \end{cases}
\end{equation}
and
\begin{equation}
    \underline{l}_n(\mathcal{I})
    =\sqrt{\delta_n^2(\mathcal{I})+\hat{y}_n^2}.
\end{equation}
The truncated Gaussian density is radially symmetric and nonincreasing around $(\hat{x}_n, \hat{y}_n)$. Therefore, for a fixed coverage radius, the covered probability is nonincreasing in the center separation $l_n$, and the required radius $r_{n,\min}^{\rm TG}(l_n)$ is nondecreasing in $l_n$. It follows that
\begin{equation}
    \underline{F}_{\rm TG}(\mathcal{I})
    =
    \sum_{n=1}^{N}c_n\left(
    \left(r_{n,\min}^{\rm TG}(\underline{l}_n(\mathcal{I}))\right)^2+d^2
    \right)
    \label{eq:tg_lower_bound}
\end{equation}
is a valid lower bound on $F_{\rm TG}(x)$ for all $x\in\mathcal{I}$. An upper bound is obtained by evaluating $F_{\rm TG}(x)$ at any feasible point in the interval, such as its midpoint. The branch-and-bound procedure repeatedly bisects the interval with the smallest lower bound, updates the incumbent feasible solution, and discards any interval whose lower bound is no smaller than the current upper bound. It terminates when the global upper--lower bound gap is below a prescribed tolerance $\epsilon_{\rm B}$, thereby producing an $\epsilon_{\rm B}$-globally optimal antenna position.

As $r_{\rm u}\rightarrow\infty$, the normalization factor in \eqref{eq:tg_normalization} approaches one and the bounded model converges to the unbounded Gaussian model.
}

\begin{figure*}[htbp]
\centering
\subfloat[\label{fig:1a}]{\includegraphics[width=0.3\textwidth]{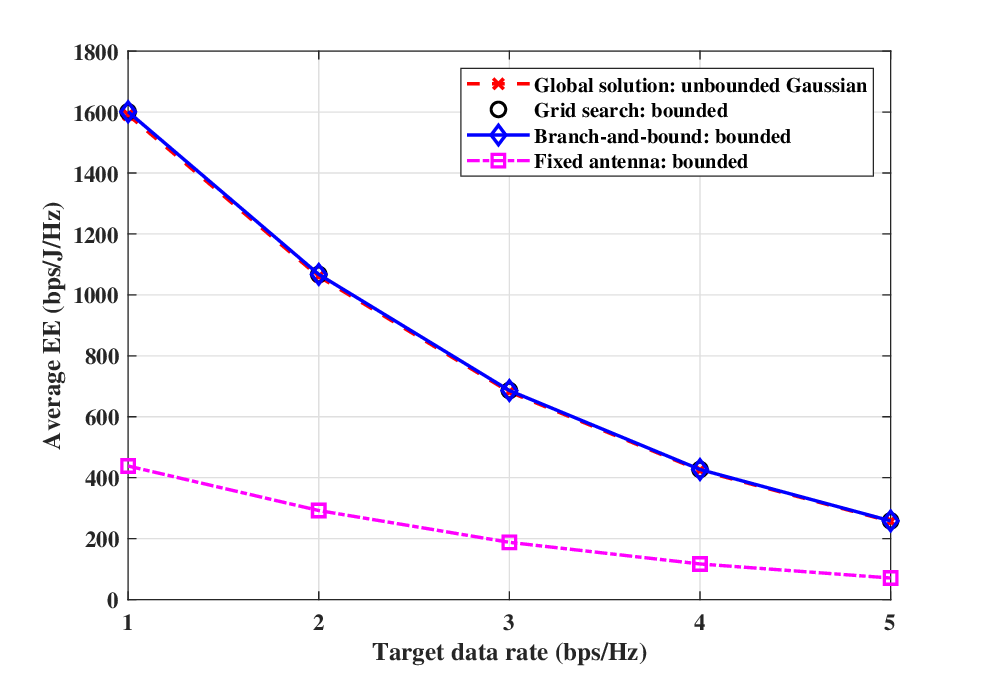}}\hfill
\subfloat[\label{fig:1b}]{\includegraphics[width=0.3\textwidth]{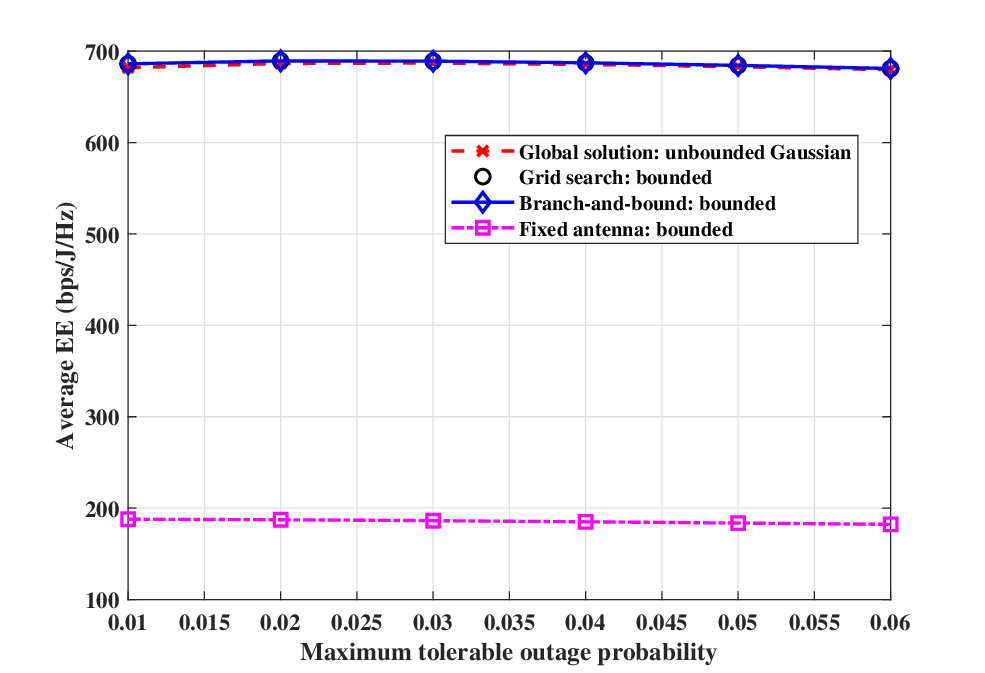}}\hfill
\subfloat[\label{fig:1c}]{\includegraphics[width=0.3\textwidth]{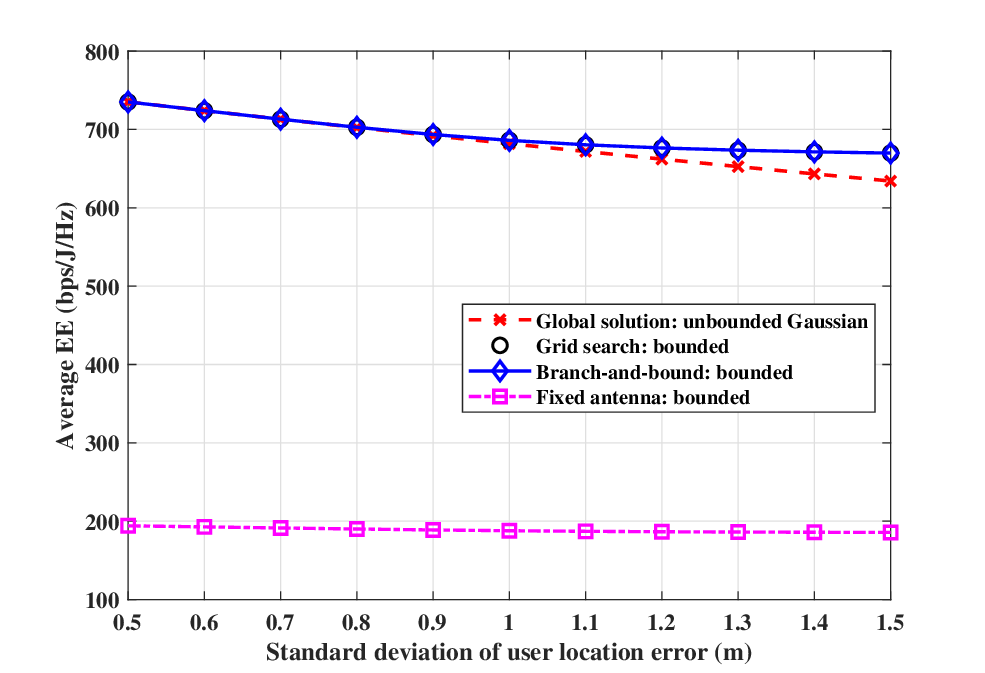}}
\caption{{\color{black}Average EE versus (a) target data rate, (b) maximum tolerable outage probability, and (c) standard deviation of the user-location error. The bounded-Gaussian schemes use $r_{\rm u}=3$~m. {For these schemes, $\sigma$ is the original Gaussian error scale.}}}
\label{fig:three_figures}
\end{figure*}


\section{Numerical Results}
Here we report numerical simulations to evaluate the proposed robust pinching-antenna designs under Gaussian user-location uncertainty. Unless otherwise specified, the system serves $N=5$ users, whose estimated locations are independently and uniformly distributed over a $120\,\mathrm{m}\times20\,\mathrm{m}$ rectangular region. The standard deviation of the location error is $\sigma=1$~m, the target spectral efficiency is $\hat{R}_n=3$~bps/Hz, and the maximum tolerable outage probability is $\varepsilon_n=0.01$, $\forall n$. The carrier frequency is 28~GHz, the aggregate bandwidth is 100~MHz, and the receiver noise power is $-94$~dBm. The waveguide has length $L=50$~m and height $d=3$~m. {\color{black}For the circularly truncated Gaussian model, the truncation radius is set to $r_{\rm u}=3$~m.} All results are averaged over 100 independent user-location realizations; {\color{black}the outage probabilities themselves are evaluated analytically or deterministically rather than through Monte Carlo rare-event counting. 

Four schemes are compared in Fig.~\ref{fig:three_figures}. The \emph{global solution: unbounded Gaussian} applies the derivative-bisection method in Section~III-B to the original unbounded Gaussian model. The \emph{branch-and-bound: bounded} scheme applies the deterministic branch-and-bound method in Section~IV to the circularly truncated Gaussian model. To verify its numerical optimality, the \emph{grid search: bounded} benchmark evaluates the same bounded-Gaussian objective over 100 uniformly spaced antenna positions in $[0,L]$. Finally, the \emph{fixed antenna: bounded} benchmark fixes the antenna at $x^{\rm Pin}=0$ and optimizes only the transmit powers under the same bounded-Gaussian uncertainty model. Hence, the comparison between the branch-and-bound and fixed-antenna schemes isolates the EE gain provided by antenna-position optimization, whereas the comparison between branch-and-bound and grid search validates the proposed bounded-Gaussian solver.

Fig.~\ref{fig:1a} plots the average EE versus the common target data rate. The EE decreases rapidly as the target rate increases for all schemes. This follows directly from the EE definition: the effective sum rate increases only linearly with $\hat{R}_n$, whereas the required transmit power contains the factor $2^{\hat{R}_n}-1$ in \eqref{Pmin} and therefore grows exponentially. The bounded branch-and-bound and grid-search curves are virtually indistinguishable over the entire rate range, numerically confirming the accuracy of the proposed branch-and-bound solution. The unbounded-Gaussian global solution also almost overlaps the two bounded optimized curves. Under the baseline setting $\sigma=1$~m and $r_{\rm u}=3$~m, the normalization factor in \eqref{eq:tg_normalization} is $1-\exp(-9/2)\approx0.989$. In contrast, the fixed-antenna benchmark exhibits substantially lower EE at every target rate, clearly demonstrating the benefit of optimizing the pinching-antenna position.

Fig.~\ref{fig:1b} shows the average EE versus the maximum tolerable outage probability. Unlike the target-rate case, the optimized EE varies only mildly over the considered range of $\varepsilon_n$. Increasing $\varepsilon_n$ has two competing effects: the effective sum rate $\sum_n(1-\varepsilon_n)\hat{R}_n$ decreases, while the reliability requirement is relaxed, which reduces the minimum coverage radius and hence the required transmit power. The new results therefore exhibit an almost flat behavior with a shallow maximum at an intermediate outage probability rather than a strictly monotonic trend. Again, the bounded branch-and-bound and grid-search curves coincide, while the unbounded-Gaussian curve remains extremely close because $r_{\rm u}=3$~m retains almost all probability mass when $\sigma=1$~m. The fixed-antenna architecture remains far below all optimized-placement schemes, confirming that antenna mobility provides a substantial EE gain throughout the considered outage-probability range.

Finally, Fig.~\ref{fig:1c} investigates the effect of the location-error standard deviation $\sigma$. The EE decreases with increasing $\sigma$ because greater location uncertainty requires a larger coverage radius to satisfy the same outage constraint, thereby increasing the required transmit power. The branch-and-bound curve once again coincides with the bounded grid-search benchmark. A particularly useful observation is that the bounded and unbounded Gaussian models are nearly identical for small $\sigma$ but progressively separate as $\sigma$ increases. This behavior follows from \eqref{eq:tg_normalization}: for fixed $r_{\rm u}$, a larger $\sigma$ places more Gaussian probability mass outside the truncation disk. Conditioning the bounded model on the disk therefore removes increasingly significant tail realizations, resulting in a smaller required coverage radius and higher EE than in the unbounded model. 
Although the fixed-antenna EE changes only slightly with $\sigma$ over the considered range, it remains substantially below the optimized schemes, demonstrating that the benefit of antenna placement persists even under relatively severe location uncertainty.
}

\section{Conclusion} 
\label{Sec:Conclusion}
This paper investigated robust joint power allocation and antenna placement for pinching-antenna systems under Gaussian location uncertainty. Using the Marcum-Q function, we characterized the outage probability and derived the minimum required transmit power. For the unbounded Gaussian model, we established the convexity of the antenna-placement objective and developed a globally optimal derivative-bisection algorithm. For the circularly truncated Gaussian model, we proposed a deterministic branch-and-bound algorithm with an \(\epsilon_B\)-global optimality guarantee. We further showed that position-dependent waveguide attenuation may destroy convexity. Numerical results demonstrated that the proposed branch-and-bound solution closely matches the grid-search benchmark and that optimized antenna placement substantially improves energy efficiency over fixed-antenna deployment. Moreover, the bounded and unbounded Gaussian models exhibit increasing performance differences as localization uncertainty grows. Future work will investigate multi-waveguide and multi-antenna architectures under location uncertainty.

\balance


\begin{thebibliography}{30}
\bibitem{Atsushi_22}
A.~Fukuda \emph{et al.}, ``Pinching antenna using a dielectric waveguide as an antenna,'' \emph{Technical Journal}, vol.~23, no.~3, pp.~5--12, Jan.~2022.
\bibitem{yang2025}
Z.~Yang \emph{et al.}, ``Pinching antennas: Principles, applications and challenges,'' \emph{IEEE Wireless Commun.}, pp.~1--10, 2025.
\bibitem{ding2024}
Z.~Ding \emph{et al.}, ``Flexible-antenna systems: A pinching-antenna perspective,'' \emph{IEEE Trans. Commun.}, vol.~73, no.~10, pp.~9236--9253, Oct.~2025.
\bibitem{pass1}
J.~Zhao, X.~Mu, K.~Cai, Y.~Zhu, and Y.~Liu, ``Waveguide division multiple access for pinching-antenna systems (PASS),'' arXiv:2502.17781, 2025.
\bibitem{Xiao25}
J.~Xiao \emph{et al.}, ``Frequency-selective modeling and analysis for OFDM-integrated wideband pinching-antenna systems,'' \emph{IEEE Wireless Commun. Lett.}, vol.~14, no.~11, pp.~3500--3504, Nov.~2025.
\bibitem{wijewardhana2025}
K.~R.~Wijewardhana \emph{et al.}, ``Wireless-fed pinching-antenna systems (Wi-PASS) for nextg wireless networks,'' \emph{IEEE Commun. Mag.}, pp.~1--7, 2026.
\bibitem{xie2025}
X.~Xie \emph{et al.}, ``A low-complexity placement design of pinching-antenna systems,'' \emph{IEEE Commun. Lett.}, vol.~29, no.~8, pp.~1784--1788, Aug.~2025.
\bibitem{wang2024}
K.~Wang \emph{et al.}, ``Antenna activation for NOMA assisted pinching-antenna systems,'' \emph{IEEE Wireless Commun. Lett.}, vol.~14, no.~5, pp.~1526--1530, May~2025.
\bibitem{hu2025}
S.~Hu \emph{et al.}, ``Sum-rate maximization for pinching antenna-assisted NOMA systems with multiple dielectric waveguides,'' arXiv:2503.10060, 2025.
\bibitem{Li25}
Y.~Li \emph{et al.}, ``Pinching antenna-aided wireless powered communication networks,'' \emph{IEEE Wireless Commun. Lett.}, vol.~15, pp.~255--259, 2026.
\bibitem{Zeng_COMML25}
M.~Zeng \emph{et al.}, ``Sum rate maximization for NOMA-assisted uplink pinching-antenna systems,'' \emph{IEEE Wireless Commun. Lett.}, vol.~15, pp.~280--284, 2026.
\bibitem{fu2025}
Y.~Fu \emph{et al.}, ``Power minimization for NOMA-assisted pinching antenna systems with multiple waveguides,'' arXiv:2503.20336, 2025.
\bibitem{Tegos_2025}
S.~A.~Tegos \emph{et al.}, ``Minimum data rate maximization for uplink pinching-antenna systems,'' \emph{IEEE Wireless Commun. Lett.}, vol.~14, no.~5, pp.~1516--1520, May~2025.
\bibitem{Zhao_TCOM25}
J.~Zhao \emph{et al.}, ``Pinching-antenna systems-enabled multi-user communications: Transmission structures and beamforming optimization,'' \emph{IEEE Trans. Commun.}, pp.~1--1, 2025.
\bibitem{zeng2025_WCM}
M.~Zeng \emph{et al.}, ``Resource allocation for pinching-antenna systems: State-of-the-art, key techniques and open issues,'' arXiv:2506.06156, 2025.
\bibitem{Xiao_COMML25}
J.~Xiao \emph{et al.}, ``Channel estimation for pinching-antenna systems (PASS),'' \emph{IEEE Commun. Lett.}, vol.~29, no.~8, pp.~1789--1793, Aug.~2025.
\bibitem{Sun_TVT26}
M.~Sun, C.~Ouyang, S.~Wu, and Y.~Liu, ``Robust beamforming for pinching-antenna systems,'' \emph{IEEE Trans. Veh. Technol.}, pp.~1--6, 2026.
\bibitem{zeng2025robust}
M.~Zeng \emph{et al.}, ``Robust resource allocation for pinching-antenna systems under user location uncertainty,'' \emph{IEEE Trans. Veh. Technol.}, pp.~1--5, 2026.
\bibitem{Muppirisetty2016}
{L.~S.~Muppirisetty, T.~Svensson, and H.~Wymeersch, ``Spatial wireless channel prediction under location uncertainty,'' \emph{IEEE Trans. Wireless Commun.}, vol.~15, no.~2, pp.~1031--1044, Feb.~2016.}
\bibitem{Kallehauge2024}
{T.~Kallehauge, \textit{et al.}, ``On the statistical relation of ultra-reliable wireless and location estimation,'' \emph{IEEE Trans. Wireless Commun.}, vol.~23, no.~8, pp.~10044--10057, Aug.~2024.}
\bibitem{Uchimura2026}
{S.~Uchimura, \textit{et al.}, ``Deterministic and stochastic optimization for robust beamfocusing against positioning errors,'' \emph{IEEE Trans. Wireless Commun.}, vol.~25, pp.~2811--2825, 2026.}
\bibitem{LiuVehicular2020}
{F.~Liu, \textit{et al.}, ``Radar-assisted predictive beamforming for vehicular links: Communication served by sensing,'' \emph{IEEE Trans. Wireless Commun.}, vol.~19, no.~11, pp.~7704--7719, Nov.~2020.}
\bibitem{Tyrovolas_IoTJ26}
D.~Tyrovolas \emph{et al.}, ``How many pinching antennas are enough?'' \emph{IEEE Internet Things J.}, vol.~13, no.~10, pp.~21994--22006, 2026.
\bibitem{Fang_TCOM21}
F.~Fang \emph{et al.}, ``Energy-efficient resource allocation for NOMA-MEC networks with imperfect CSI,'' \emph{IEEE Trans. Commun.}, vol.~69, no.~5, pp.~3436--3449, May~2021.
\bibitem{Zhang_TVT17}
Y.~Zhang \emph{et al.}, ``Energy-efficient transmission design in non-orthogonal multiple access,'' \emph{IEEE Trans. Veh. Technol.}, vol.~66, no.~3, pp.~2852--2857, Mar.~2017.
\bibitem{Zeng_TVT19}
M.~Zeng \emph{et al.}, ``Spectral- and energy-efficient resource allocation for multicarrier uplink NOMA systems,'' \emph{IEEE Trans. Veh. Technol.}, vol.~68, no.~9, pp.~9293--9296, 2019.
\bibitem{Marvin}
M.~K.~Simon, \emph{Probability Distributions Involving Gaussian Random Variables}. Springer, 2002.
\bibitem{Pratt_IEEEProc68}
W.~Pratt, ``Partial differentials of Marcum's Q function,'' \emph{Proc. IEEE}, vol.~56, no.~7, pp.~1220--1221, 1968.
\bibitem{segura2021monotonicity}
J.~Segura, ``Monotonicity properties for ratios and products of modified Bessel functions and sharp trigonometric bounds,'' \emph{Results Math.}, vol.~76, no.~4, Art.~no.~221, 2021.
\bibitem{tyrovolas2025}
D.~Tyrovolas \emph{et al.}, ``Performance analysis of pinching-antenna systems,'' \emph{IEEE Trans. Cogn. Commun. Netw.}, vol.~12, pp.~590--601, 2026.
\end{thebibliography}
\end{document}